\documentclass[journal, nocompress]{IEEEtran}

\usepackage{amsmath}
\usepackage{physics}
\usepackage{array}
\usepackage{cite}
\usepackage{makecell}

\usepackage[caption = false, font = footnotesize, labelfont = sf, textfont = sf]{subfig}

\usepackage{floatrow}
\usepackage{multirow}
\usepackage{tabu}
\usepackage{tabularx}
\newcolumntype{C}[1]{>{\centering\arraybackslash}p{#1}}
\usepackage{hhline}
\usepackage[pdftex]{graphicx}
\DeclareGraphicsExtensions{.pdf,.jpeg,.png,.bmp,.jpg,.tif}
\usepackage{stfloats}

\usepackage{float}

\def\colorModel{hsb} 

\newcommand\ColCell[1]{
  \pgfmathparse{#1<50?1:0}  
    \ifnum\pgfmathresult=0\relax\color{white}\fi
  \pgfmathsetmacro\compA{0}      
  \pgfmathsetmacro\compB{#1/100} 
  \pgfmathsetmacro\compC{1}      
  \edef\x{\noexpand\centering\noexpand\cellcolor[\colorModel]{\compA,\compB,\compC}}\x #1
  } 
\newcolumntype{E}{>{\collectcell\ColCell}m{0.4cm}<{\endcollectcell}}  

\begin{document}

\title{DFR-TSD: A Deep Learning Based Framework for Robust Traffic Sign Detection Under Challenging Weather Conditions}

\author{
Sabbir Ahmed,
Uday Kamal, and, 
Md. Kamrul Hasan\IEEEauthorrefmark{1}

\thanks{All authors are with the Department of Electrical and Electronic Engineering, Bangladesh University of Engineering and Technology, Dhaka-1205, Bangladesh.}
\thanks{E-mail:  \IEEEauthorrefmark{1}khasan@eee.buet.ac.bd}
}

\maketitle

\begin{abstract}
Robust traffic sign detection and recognition (TSDR) is of paramount importance for the successful realization of autonomous vehicle technology. The importance of this task has led to a vast amount of research efforts and many promising methods have been proposed in the existing literature. However, the SOTA (SOTA) methods have been evaluated on clean and challenge-free datasets and overlooked the performance deterioration associated with different challenging conditions (CCs) that obscure the traffic images captured in the wild. In this paper, we look at the TSDR problem under CCs and focus on the performance degradation associated with them. To overcome this, we propose a Convolutional Neural Network (CNN) based TSDR framework with prior enhancement. Our modular approach consists of a CNN-based challenge classifier, Enhance-Net, an encoder-decoder CNN architecture for image enhancement, and two separate CNN architectures for sign-detection and classification. We propose a novel training pipeline for Enhance-Net that focuses on the enhancement of the traffic sign regions (instead of the whole image) in the challenging images subject to their accurate detection. We used CURE-TSD dataset consisting of traffic videos captured under different CCs to evaluate the efficacy of our approach. We experimentally show that our method obtains an overall precision and recall of 91.1\% and 70.71\% that is 7.58\% and 35.90\% improvement in precision and recall, respectively, compared to the current benchmark. Furthermore, we compare our approach with SOTA object detection networks, Faster-RCNN and R-FCN, and show that our approach outperforms them by a large margin. 
\end{abstract}


\section{Introduction}
\IEEEPARstart{T}{raffic} sign detection and recognition play a crucial part in driver assistance systems and autonomous vehicle technology. One of the major prerequisites of safe and widespread implementation of this technology is a TSDR algorithm that is not only accurate but also robust and reliable in a variety of real-world scenarios. However, in addition to the large variation among the traffic signs to detect, the traffic images that are captured in the wild are not ideal and often obscured by different adverse weather conditions and motion artifacts that substantially increase the difficulty level of this task. 


Being a challenging research problem, several studies have been carried out on TSDR. A detailed overview of these studies can be found in \cite{mogelmose2012vision} and \cite{gudigar2016review}. The whole task of TSDR can be subdivided into two independent tasks: Traffic Sign Detection (TSD) and Traffic Sign Recognition (TSR). Traditional research methodologies of TSD  mostly rely on manual feature extraction of various attributes such as geometrical shapes, edge detection, and color information. Color-based approach mostly comprises of threshold-based segmentation of traffic sign region in a particular color space such as Hue-Saturation-Intensity (HSI) \cite{xu2009robust}, Hue-Chroma-Luminance (HCL) \cite{khan2011image} and others \cite{creusen2010color}. 
However, one major drawback of these color-based approaches is that these are highly susceptible to the change in illumination that can frequently occur in real-world scenarios \cite{temel_2}. To overcome this challenge, shape-based approaches have been extensively used in the existing literature that comprise of Canny-Edge detection \cite{canny1987computational}, Histogram-Oriented Gradients (HOG) \cite{creusen2010color}, Haar-Wavelet features \cite{baro2009traffic} and Fast Fourier Transform (FFT) \cite{jimenez2008traffic}. However, size and scale variation, disorientation and  occlusions of traffic sign regions due to the motion artifacts during real-time video feed hinder the practical application of these approaches. On the other hand, the methodologies for TSR utilize color and/or shape-based features to train classifiers such as Random Forest \cite{zaklouta2014real} and Support Vector Machine (SVM) \cite{creusen2010color}. However, there is no comprehensive method to identify the best features set and the best classifier for TSR.

Recent advancement of deep learning (DL) in different computer vision tasks such as image recognition \cite{krizhevsky2012ImageNet}, image segmentation \cite{chen2014semantic} and object detection problems \cite{girshick2014rich} has led to a large scale adoption of these algorithms for TSDR \cite{conv1, conv2, frcnn_mobilenet}. Lee \textit{\textit{et al.}} have used a custom designed CNN to simultaneously detect and estimate the boundary of the traffic sign regions\cite{lee}. Compared to the objects that frequently appear in the existing object detection datasets, traffic sign regions are very small and thus have a very small region of interest (ROI) to background ratio. To address this challenge, Yuan \textit{et al.} have proposed a CNN-based multi-resolution feature fusion architecture\cite{vssa}. They have also proposed a vertical spatial sequence attention (VSSA) module to gain more context information for better detection performance. A major advantage of DL-based methods is that they are completely data-driven that does not require any manual feature engineering. In order to facilitate the development and evaluation of these algorithms, a number of datasets have been introduced such as GTSDB\cite{houben2013detection}, LISA\cite{lisa}, BelgiumTS\cite{timofte2014multi} and TT100K\cite{timofte2014multi}. However, neither these methods nor these datasets have considered sufficient types and levels of CCs that can frequently arise during the capture of the traffic image \cite{temel_2}.  


In order to overcome this shortcoming of the existing datasets, Temel \textit{et al.} have introduced a video dataset that has considered both the Challenging Unreal (stimulated) and Real Environments for Traffic Sign Detection (CURE-TSD)\cite{temel2017cure-tsd}. This dataset comprises of different CCs that can commonly arise in practical scenarios with more than 1.7 million frames that makes it by far the largest dataset available for traffic sign detection research. And in \cite{temel_2}, Temel \textit{et al.} have analyzed the effect of various CCs on the performance of benchmark algorithms. 
They have presented the two benchmarks provided by the top two winners of the IEEE Video and Image Processing(VIP) Cup 2017 and showed that the CCs can reduce the $F_{2}$ score by 65\%. Such significant performance degradation undoubtedly calls for special attention to these challenges.  

In \cite{uday}, Kamal \textit{et al.} have proposed a deep CNN-based modular approach that achieves SOTA performance on this dataset under challenge free conditions. The authors have also highlighted their performance degradation associated with the CCs. A straightforward solution to alleviate this problem is to use an image enhancer that retains proper color and shape information corrupted by various CCs. Many researches have been conducted on image enhancement task for the purpose of better visibility. However, in most of these studies, only one challenge case is considered \cite{derain, derain_2, dehazing}. In very few studies, two challenge cases are considered \cite{rain_and_snow,rain_and_snow_1}. Therefore, the development of a robust image enhancer that can address multiple challenge cases with varying severity is one of the most important considerations for accurate sign detection in real-world scenarios. 


To alleviate this problem, we propose Enhance-Net, a deep CNN-based image enhancer, that performs a prior enhancement of these traffic images. 
In \cite{temel_2}, Temel \textit{et al.} has stated that all of the twelve different types of CCs present in the dataset need not be addressed at a time and also highlight the fact that benchmark algorithms performance is more vulnerable in challenging weather conditions such as Rain, Snow, and Haze. In addition, Lens blur and Dirty Lens are two frequently occurring CCs in the real world scenarios. Therefore, in this work, we highlight the focus of our approach on five different CCs: Rain, Snow, Haze, Dirty lens, and Lens blur associated with five different levels of severity. Due to the different nature of these challenges, training a single network for all these challenges may result in sub-optimal performance. Therefore, we use five different Enhance-Nets trained with a single type of challenge at a time to ensure the best possible enhancement for each of the CCs. We incorporate the modular TSDR approach proposed in \cite{uday} for sign detection and classification from the enhanced traffic images. Therefore, our approach consists of four separate modules- challenge classifier, enhancement blocks, sign detector, and sign classifier. The challenge classifier classifies the challenge present in the image and forwards it to the appropriate enhancement block that enhances the corresponding challenging image. The enhanced image is then passed through the sign detector that localizes the traffic signs and the sign classifier assigns them to the appropriate sign class. The challenge classifier that classifies these challenges can also detect the challenge-free cases and therefore our approach is robust in challenge-free conditions as well. 


In addition, we propose a novel training pipeline for Enhance-Net that emphasizes the enhancement of traffic sign regions in contrast to the existing training methods that perform enhancement of the whole image. Although these existing methods may lead to better visibility of the overall image, we experimentally show that it offers very limited improvement in traffic sign detection performance. Because an overall enhancement of the image does not necessarily ensure equally enhanced traffic sign regions. To address this issue, our proposed method incorporates the calculation of the loss function only on the traffic sign regions. As the end goal of the approach is to ensure better TSDR, the purpose of the enhancement blocks is not only to ensure the better visibility of the sign regions but also their better detection by the sign detector. In addition, Ledig \textit{et al.} have demonstrated that pixel-wise MSE-loss optimization alone cannot ensure the proper reconstruction of high-frequency contents and, therefore, proposed an MSE-loss in the feature domain of a pretrained deep CNN, termed as content loss \cite{srgan}. In TSDR, such high-frequency contents can be crucial for the subsequent sign detection and classification module. Therefore, to ensure the best possible performance of the Enhance-Nets, we propose a novel loss function that combines the pixel-domain and feature-domain reconstruction loss as well as the detection loss of this enhanced sign regions. With proper optimization of this loss function and our training pipeline that ensures both the enhancement and successful detection of the sign regions, we are able to achieve a substantial improvement of the current benchmark on the challenging part of the dataset.  

In this work, we extend the benchmark result of Kamal \textit{et al.} on the challenge-free data of the CURE-TSD dataset by incorporating the results of our approach on five different CCs with five different levels of severity. To demonstrate the efficacy of our proposed modular solution, we compare its performance with two deep CNN-based end-to-end approaches-- Faster R-CNN Inception Resnet V2 and R-FCN Resnet 101 \cite{arcos2018evaluation} that are trained with both challenge free and challenging data. The experimental results reveal that our proposed solution not only improves the current SOTA result on the challenging data but also outperforms the aforementioned methods by a large margin. 

In the following sections of this paper, we first provide a detailed description of each part of our proposed method as well as their underlying motivation (Section II), a brief description of the CURE-TSD dataset along with the training procedure of the networks (Section III), a thorough discussion of the experimental results obtained after evaluating our method (Section IV) and finally, some concluding remarks of this work. (Section V).

\section{Methodology}
\subsection{Motivation}\label{motivation}
Traffic sign detection is often viewed as an image segmentation and recognition problem. The authors of SegU-Net \cite{uday} has reported SOTA performance on CURE-TSD dataset. While their performance on the challenge-free dataset is appealing, they show that incorporation of the challenging dataset significantly deteriorates the performance. This is due to the fact that the adversaries present in those images obscure the color and shape information of small traffic sign regions in particular. Therefore, an overall approach that is robust to the adverse effect of these CCs is of utmost importance for practical application\cite{frcnn_mobilenet}.

\begin{figure*}[!t]
	\centering
	\includegraphics[width=7in,
	height=2.2in]{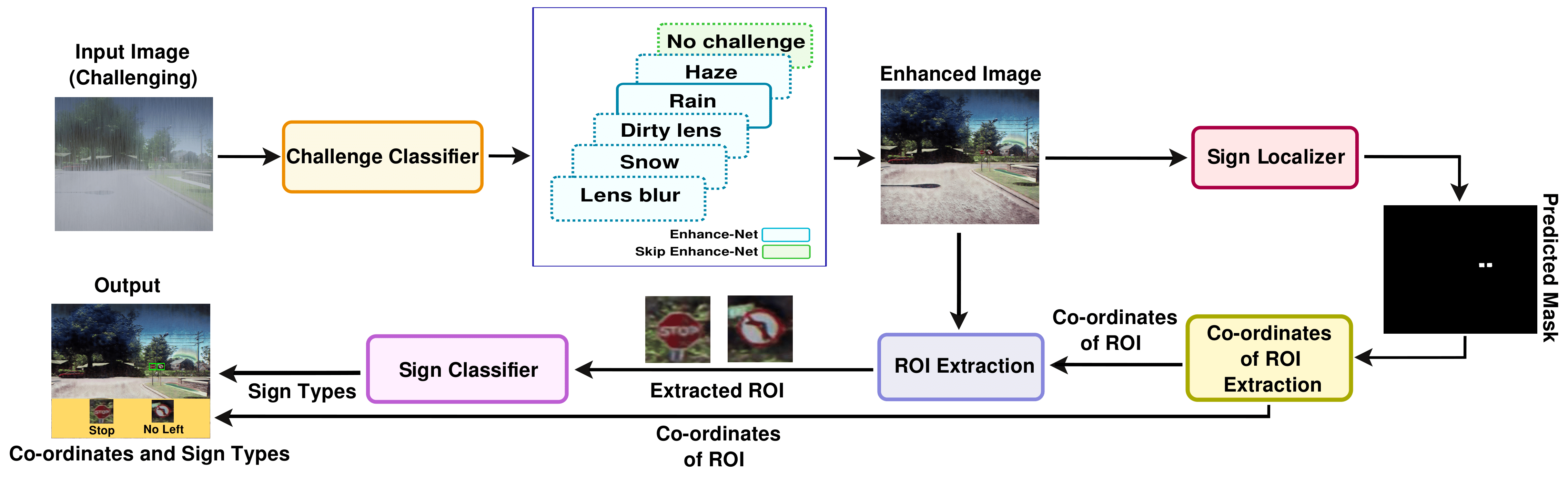}
	\caption{Outline of the proposed pipeline for robust traffic sign detection and classification.}
	\label{fig_model}
\end{figure*}

While many promising CNN-based approaches have been proposed in the literature for traffic sign detection and classification, none of them considers such CCs. As a result, the performance of these methods are severely vulnerable to the quality of the captured image. One straightforward approach can be re-training the networks on the challenging datasets. However, training of these large scale models is not only a tremendous task but also it does not guarantee superior performance in these conditions, which is experimentally demonstrated in the later section. Therefore, we propose a solution to this problem that is completely modular in structure that allows others to incorporate their sign detection model already trained on the challenge-free condition and achieve significantly better performance even in the CCs without requiring any modification to their architectures or trained weights. Although it is possible to work on a general-purpose architecture to handle all kinds of challenges, this would result in sub-optimal performance. This is due to the fact that the signal generation model for different kinds of challenges are significantly different in nature. For example, blur is a convolutive noise while rain is additive noise. Therefore, to ensure optimum performance, different networks can be constructed for different challenges. However, this approach requires careful consideration of an extensively large design space that is not feasible. To alleviate this problem, we utilize a challenge classifier module that enables us to utilize the same challenge enhancer architecture for different challenges. We train the same enhancement module for different challenges independently and it is the challenge classifier that guides the input image to the appropriate enhancement module.  

While most of the image enhancement works involve the enhancement of the whole image, this is not necessarily required in our case. For successful traffic sign detection, it is only the sign regions that require special attention during the enhancement operation. To validate this proposition, we perform extensive experiments (discussed in the later section) which demonstrate the fact that an overall enhanced image indeed does not introduce any significant performance gain. Therefore, we propose a novel training scheme and a carefully designed loss function that makes the enhancement module focus on the traffic sign regions during training. We also emphasize on the fact that enhancement of these sign regions should be done in such a way that works in favor of the subsequent sign detection module. For instance, an enhanced image with better visibility can loose some spectral or inherent information during the reconstruction process that can be crucial for its successful detection by the sign-detector. Therefore, to ensure the optimum performance of the detection module on these enhanced images, our proposed method also incorporates the sign detection loss during the training of the enhancement blocks so that an overall minimization of the loss ensures better quality of the sign regions as well as their successful detection. 

Overall, our proposed method consists of four separate modules. The first module, Challenge Classifier, detects the type of challenge present in the traffic image. The second module, Enhancement Block, performs the required enhancement for different challenges. 
Finally, the third and the fourth module consist of a Sign Detection and a Classification block, respectively.  We use the architectures proposed by in \cite{uday} that outperform other SOTA object detection models such as Faster-R CNN and R-FCN \cite{arcos2018evaluation} by a large margin. The outline of our proposed method is as follows (see Fig. \ref{fig_model}):

\begin{enumerate}
	\item First, the challenge classifier detects the type of challenge present in the traffic image.
	\item If the detected challenge type falls within 5 categories (Rain,  Snow,  Haze,  Dirty lens and Lens blur), the image is passed to the corresponding enhancement block.
    \item The enhanced image is finally passed to SegU-Net and sign classifier for detection and classification of traffic signs.
	\item  If the detected challenge type is `No challenge', the image bypasses the enhancement blocks and is passed directly to SegU-Net and the sign classifier.
	
\end{enumerate}

\subsection{Challenge Classifier}
The challenge classifier is implemented by utilizing transfer learning on VGG16\cite{vgg} CNN architecture pretrained on the ImageNet dataset. The network consists of two stages: feature extraction stage and classification stage. The feature extractor compresses the information present in the image to a lower-dimensional feature space. The subsequent stage uses these features to perform the desired classification.

\paragraph{Feature Extractor} The feature extractor consists of thirteen convolution blocks, each with $3\times 3$ convolution kernel, a batch normalization layer and a ReLU activation layer. For downsampling purpose, the max-pooling operation is used. Finally, a global average-pooling operation is performed in the final stage to compress the features further.

\paragraph{Classification} Classification stage has a fully connected layer - that has an output size of 6 (five different types of challenge and one no-challenge condition) with softmax activation. This stage utilizes the extracted features for classification.

\subsection{Enhancement Blocks}
We use five enhancement blocks for addressing the five different challenges. Each of these blocks consists of the same CNN-based network architecture that is trained on five different types of challenge separately and independently. This provides ease of adding more enhancement blocks to address other challenges as well. 


\begin{figure*}[!t]
	\centering
	\includegraphics[width=7in, height=2.2in]{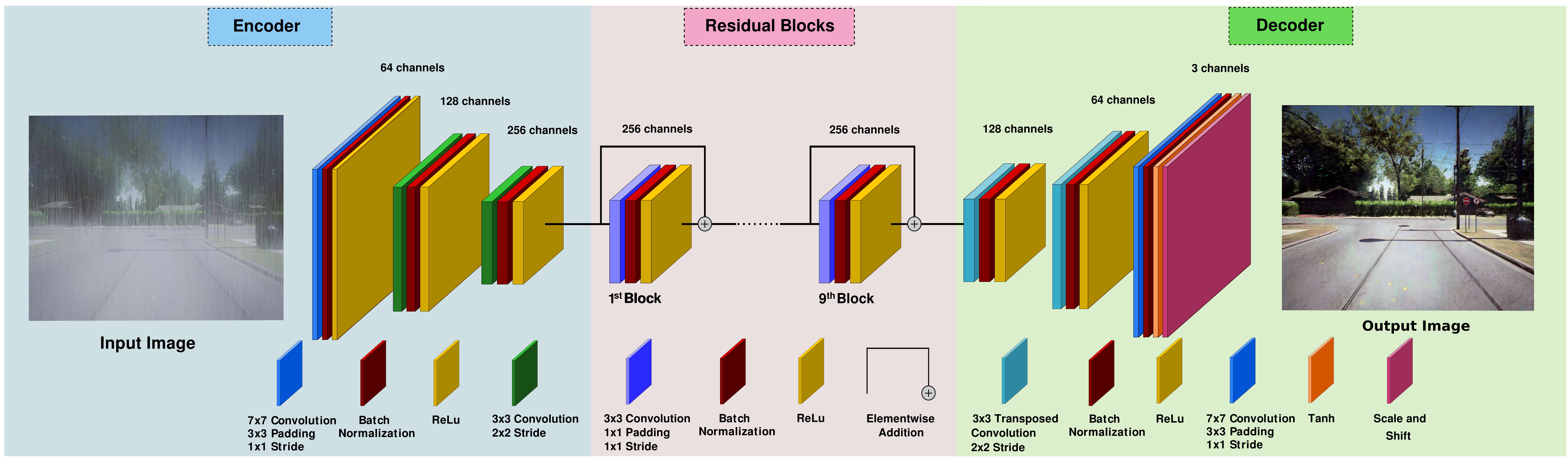}
	\caption{Architecture of Enhance-Net which consists of three blocks: Encoder block, Residual blocks and Decoder block. The downsampling operation in Encoder block is performed by $2\times 2$ strided convolution operation. Element-wise addition operation has been adopted in residual blocks. Transposed convolution operation in the decoder blocks regenerates the enhanced version of the image from the compressed feature space. Finally, the Shift and Scale layer converts the output into pixel domain.}
	\label{fig_architecture}
\end{figure*}
 
\subsubsection{CNN Enhancement Block} The CNN enhancement block is made of a fully CNN architecture \cite{long2015fully}, which is inspired by the architecture of the generator part of DeblurGAN\cite{dgan}, a SOTA CNN-based image deblurring method. The network is designed using an encoder (downsampling stage), nine residual blocks (with identity skip connection) followed by a decoder (upsampling stage). 

\paragraph{Encoder} The encoder stage has one convolution block with  $7\times 7$ convolution kernels, an instance normalization and a ReLU activation layer followed by two convolutional blocks, each with $3\times 3$ convolution kernels, an instance normalization, and a ReLU activation layer. The number of kernels for convolutional blocks are 64, 128, and 256 and with a stride of $1\times 1$, $2\times 2$, and $2\times 2$, respectively. The encoder encodes the image to the latent feature maps that are further enhanced by the residual blocks.

\paragraph{Residual Block} Each Residual block is composed of two convolution layers each with $3\times 3$ convolution filters, an instance normalization layer and a ReLU activation layer with shortcut skip connection. In this proposed architecture, 9 residual blocks are used. Residual blocks were first used in ResNet\cite{resnet}, where the authors empirically demonstrate that in deeper networks, residual block results in faster convergence and better loss function minimization. 

\paragraph{Decoder} At each step of decoding, the feature maps generated by the residual blocks are  up-sampled  by  using two transposed-convolution blocks, each with $3\times 3$  transposed-convolution with strides of $2\times 2$, an instance normalization layer, and a ReLU activation layer. Finally, the decoding stage involves a convolution block that has $7\times 7$  sized kernels with Hyperbolic Tangent(Tanh) activation. The whole architecture is presented in Fig. \ref{fig_architecture}.


\section{Training}
In this section, we discuss the details of our training procedure. We also discuss about the dataset briefly. All of our experiments are performed in a hardware environment that includes a Intel  Core-i7  7700K,  4.50  GHz  CPU  and  Nvidia GeForce GTX 1080 Ti (11 GB Memory) GPU. All of the necessary codes are written in Python and we used Pytorch DL library to implement the neural networks.
\subsection{Dataset}\label{dataset}
CURE-TSD dataset \cite{temel2017cure-tsd} is by far the largest video dataset that emulates the practical scenarios with case of different challenging and challenge-free conditions for traffic sign detection.

\begin{table}[h]
\begin{tabular}{|C{2.75cm}||C{1cm}|C{1.4cm}|C{1cm}|}
\hline
\multirow{2}{*}{Challenge} & \multicolumn{3}{C{3.6cm}|}{Number of Videos Per Level} \\
\cline{2-4}
& Train & Validation & Test\\
\hline
Haze &  29 & 5 & 15\\
& & &\\
\hline
Rain, Snow, Dirty lens, Lens blur &  58 & 10 & 30\\

\hline     
\end{tabular}
\caption{Data distribution of CURE-TSD.}
\label{tab_data}
\end{table}



\begin{figure*}[!t]
	\centering
	\includegraphics[width=7in]{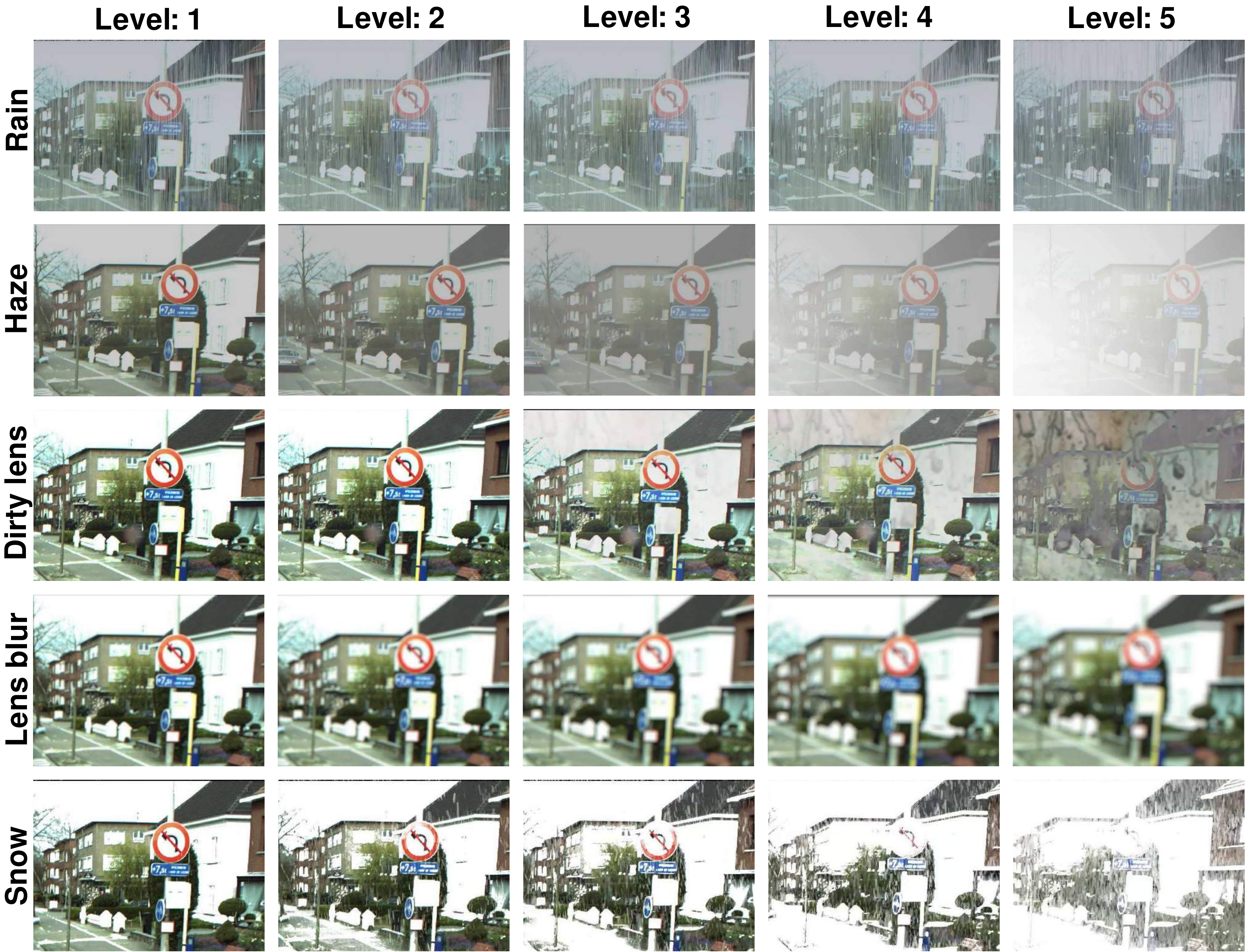}
	\caption{Selected challenges corresponding to a sample frame with five increasing levels of severity.}
	\label{fig_challenges}
\end{figure*}

This dataset comprises of videos of twelve different CCs as well as videos of challenge-free condition. The details of the dataset is are discussed elaborately in \cite{temel2017cure-tsd,uday}. We split the dataset into train, validation and test partitions according to the rules of IEEE Video and Image Processing (VIP) cup 2017. Table \ref{tab_data} contains a brief outline of this partitioning scheme. We perform all of our experiments on the training and validation dataset and evaluate the models performance on the holdout test dataset. In this work, we address and enhance the images of five different CCs (Rain, Snow, Haze, Dirty lens, and Lens blur) that can arise frequently in real-world applications. Some of the challenges such as, Codec Error, Shadow, and Exposure are conditions that might arise due to hardware malfunctions and technical faults. These are not expected to occur due to outside environment and therefore can be controlled. Thus we direct our focus on CCs that can arise frequently due to the variation of outside environment. The nature of these challenges associated with their varying level of difficulties are presented in Fig. \ref{fig_challenges}. It is worthwhile to mention that, for Haze, CURE-TSD dataset does not have any simulated environment data, as a result, the volume of data for Haze is halved compared to other other CCs.

\subsection{Challenge Classifier} 
The frames in the CURE-TSD dataset are of $1236 \times 1628$ resolution. Using this large size of image in the challenge detection stage is a computationally expensive procedure. Moreover, the impact of weather is present in the image as a global feature. As the main focus of the challenge classifier is to identify the type of challenge, downsampling operation is more likely to strengthen this global feature. Therefore, we resized all the images of selected challenges to $512\times512$ pixels and using them we train our challenge classifier for 6 classes-- 5 for the selected challenges and 1 for the challenge-free type. We used Categorical Cross-Entropy as the loss function and used Adam \cite{kingma2014adam} to optimize our network during training. The initial learning rate was set to $10^{-3}$ and a learning rate schedule was used to decrease it by a factor of 0.5 if the validation score did not improve for 3 epochs. Finally, using these specifications, we trained our network for 20 epochs.

\subsection{CNN Enhancement Block}
For training each CNN enhancement block, first, we extract all the frames from the training video sequences that contain traffic signs of a particular challenge type. Each difficulty level has 29400 number of frames which gives a total of $29400\times 5 = 147000$ frames, each with a frame size of $1236\times 1628$ pixels. Next, we crop random patches of size $1024\times 1024$ from the frames such that it contains the traffic signs. Due to hardware constraints, we use a batch size of 1 for training. Because of such a small batch size, we perform gradient accumulation. We accumulate the gradients obtained for 5 consecutive batches and update the weights ensuring that the 5 consecutive batches are from 5 different levels of a challenge. 

We trained our CNN enhancement block using these images for 35 epochs using an initial learning rate of $10^{-3}$. We used Adam \cite{kingma2014adam} as our optimizer function. A learning rate schedule was used to decrease the learning rate by a factor of 0.5 if the validation score did not improve after 3 consecutive epochs. 

\subsubsection{Loss Function}
 As an image enhancement problem in the context of traffic sign detection, the goal of our CNN enhancement block is to enhance the challenging images so that SegU-Net \cite{uday}, already trained on challenge-free dataset, can achieve elevated performance even in the CCs. However, as we mentioned earlier, this enhancement of the image should be performed in such a way that ensures SegU-Net's successful localization of the traffic signs from these enhanced images. We achieve this by incorporating the sign detection module in training the enhancement blocks. 
 Due to our modular approach, it is important to keep the performance of the detection block in the challenge-free dataset unaffected. Therefore, in all our experiments, we use SegU-Net pre-trained on the challenge-free dataset and we keep this weight unaltered. Overall, we train the image enhancer primarily focusing on the traffic sign regions but at the same time use the detection block to constrain the model to improve traffic sign detection performance. We define the proposed loss function as

\begin{equation}
	\label{eqn_mod_loss}
	L_{total} = L_{enhance(sign)} + \lambda_{1}\cdot L_{content(sign)} + \lambda_{2}\cdot L_{localizer},
\end{equation}
where $L_{enhance(sign)}$ is the pixel-level loss across sign regions, $L_{content(sign)}$ is the loss across sign regions in the feature space and $L_{localizer}$ is the loss associated with the detection block, and $\lambda_{1}$, $\lambda_{2}$ are hyperparameters acting as coupling factors that control the combination of the losses altogether. 
 


\begin{figure*}[!t]
	\centering
	\includegraphics[width=6.5in, height=3in]{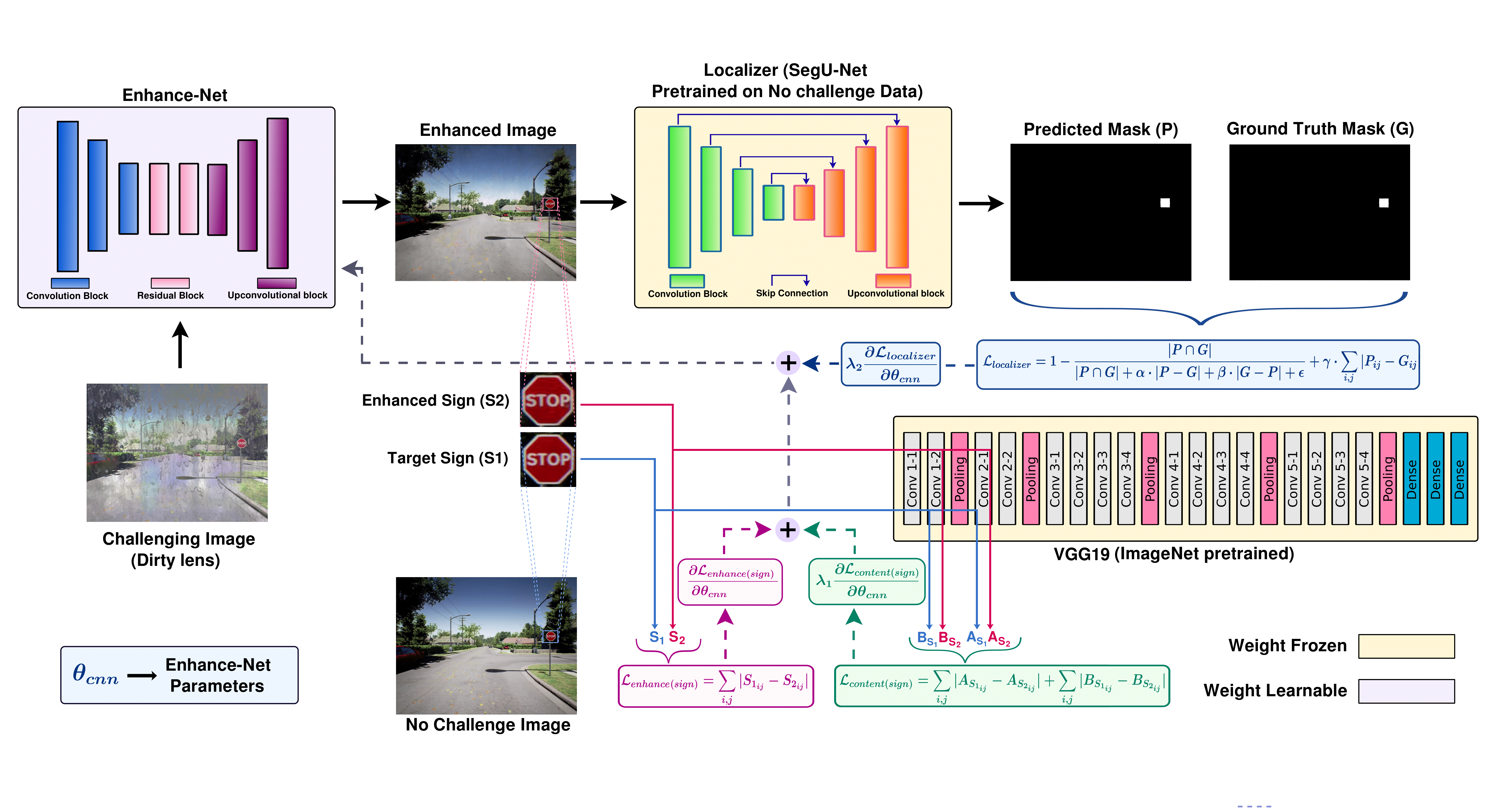}
	\caption{Training pipeline of the proposed method.}
	\label{final_images_altogether}
\end{figure*}

As overall enhancement of the challenging image may not lead to optimally enhanced traffic sign regions, we introduce a modification to the loss calculation that makes the model emphasize on the traffic sign regions only rather than the whole image. Therefore, unlike existing methods, only the sign regions contribute to the reconstruction loss in our approach. Our reconstruction loss is calculated with Mean Absolute Error (MAE) between the reconstructed and target traffic sign regions. Our choice of MAE instead of Mean Square Error (MSE) is inspired by \cite{l2} where the authors have demonstrated that using MSE loss function yields blurry reconstructed images. This loss function is not suitable for our purpose, since the blurry effect can obscure the fine details present in the low-resolution traffic signs present in the images. On the other hand, Zhao et. al \cite{l1_better} have demonstrated that using MAE as the loss function leads to better image reconstruction compared to MSE. Therefore, MAE is a better choice as a reconstruction loss for our approach. Another insight into the problem is that some challenges obscure the high-frequency components that accounts for the sharpness and fine-details of the images and thereby, adversely affect the performance of the sign detection algorithms. This suggests that alongside proper color and shape information, the high-frequency details need to be recovered as well. To meet this demand, we incorporate another loss that ensures better recovery of such details. It is widely accepted that edges, contours and other details present in an image are captured by the shallow layers of a CNN \cite{cnn_features}. Therefore, a lower level intermediate layer of a CNN pretrained on ImageNet dataset can act as the feature extractor that corresponds to the edges, contours and low-level details of the image. We adopt a VGG19 network pretrained on the ImageNet dataset and we extract features of the reconstructed image and target image traffic sign regions from its intermediate layers. Then we minimize the MAE between these reconstructed traffic sign features and target traffic sign features. This minimization ensures that the reconstructed image has recovered the necessary fine details \cite{edge_loss}. The overall calculation of loss and training scheme of the CNN enhancement block is illustrated in Figure \ref{final_images_altogether}. If $R$ and $T$ are the reconstructed and target traffic sign regions, and H, W, and C represent height, width, channels of the traffic sign regions, respectively, we define $L_{enhancement(sign)}$ and $L_{enhancement(content)}$ as

%
\begin{equation}
    L_{enhancement(sign)} = \dfrac{1}{H\times W \times C}\sum_{k=0}^{C}\sum_{j=0}^{W}\sum_{i=0}^{H}|R_{ijk} - T_{ijk}|,
\end{equation}

\begin{equation}
    L_{content(sign)} = \dfrac{1}{H\times W \times C}\sum_{k=0}^{C}\sum_{j=0}^{W}\sum_{i=0}^{H}|\phi(R)_{ijk}-\phi(T)_{ijk}|,
\end{equation}
where $\phi(R)$ and $\phi(T)$ represent feature extracted by the VGG19 network from the traffic sign region of the reconstructed and target image, respectively.


We incorporate the sign detection module, i.e. Seg-UNet, to train the enhancement block by passing the enhanced image to the detector for generating a binary segmentation mask. It is then used along with the ground truth mask to calculate the modified Tversky loss with L1 constrain that is used in \cite{uday} to train the detector on challenge-free dataset. This loss is defined as

\begin{align}
	\label{eqn_localizer}
	L_{localizer} = 1 &- \dfrac{|P \cap G|}{|P \cap G| + \alpha\cdot |P - G| + \beta\cdot |G - P| + \epsilon}\nonumber\\
	&+ \gamma\cdot\sum_{i,j}|P_{ij}-G_{ij}|,
\end{align}
where $P$ and $G$ are the set of predicted mask and ground truth binary labels, respectively. To avoid numerical instability a small safety factor  $\epsilon$ is added. To weight false positives and false negatives, $\alpha$ and $\beta$ are used as parameters where $0 \leq \alpha, ~\beta \leq 1$ with $\alpha + \beta = 1$. Here $\gamma$ is another hyperparameter that acts as a coupling factor. We use the value of $\alpha$, $\beta$ and $\gamma$ as mentioned in \cite{uday}. 

Furthermore, to get the optimum values of $\lambda_{1}$ and $\lambda_{2}$, extensive experiments are carried out and based on the results achieved on validation data, a different set of coupling factors values are obtained for different challenges that are given in Table \ref{table_gamma}.

\begin{table}[h]
\begin{tabular}{|C{2cm}||C{1.5cm}|C{1.5cm}|}
\hline
\multirow{2}{*}{Challenge} & \multicolumn{2}{C{3cm}|}{Optimum Coupling Factors} \\
\cline{2-3} 
& $\lambda_{1}$ & $\lambda_{2}$ \\
\hline
Rain & 5 & 5 \\
\hline
Snow & 1 & 10 \\
\hline
Dirty lens & 10 & 10 \\
\hline
Lens blur & 0.5 & 1 \\
\hline
Haze & 0.5 & 5 \\
\hline
\end{tabular}
\caption{Optimum Coupling Factor values for each challenge.}
\label{table_gamma}
\end{table}

\section{Experimental Results}
In this section, we demonstrate the efficacy of our proposed method on the CURE-TSD dataset and also present results of ablation study to signify the importance of each module of our architecture. Furthermore, comparative results are also presented with two SOTA deep CNN-based end-to-end trained approaches to show the superiority of our modular approach.


\subsection{Performance metrics}
We evaluate the performance of our proposed method based on the following metrics:
\begin{itemize}
	\item True Positive (TP) = estimated sign region has at least 50\% overlap with ground truth sign region with sign type being correctly classified.
	\item False Positive (FP) = estimated sign region has no overlapping ground truth sign region or sign type is incorrectly identified.
	\item True Negative (TN) = no identification of traffic sign in non sign regions.
	\item False Negative (FN) = no identification of traffic sign in sign regions.
\end{itemize}
Using these metrics, two scores are defined that summarize the performance of the model:

\begin{itemize}
	\setlength\itemsep{1ex}
	\item Precision ($Pr$) = $\dfrac{\textrm{TP}}{\textrm{TP + FP}}$,
	\item Recall ($Rc$) = $\dfrac{\textrm{TP}}{\textrm{TP + FN}}$.
\end{itemize} 
%



%

%

\subsection{Performance of the Challenge Classifier}
Performance of the challenge classifier plays a vital role in our proposed method. Therefore, we first compare the performance of different SOTA CNN architectures for our challenge classifier module and show them in Table \ref{classifier_result}. From this table, we can see that VGG16 as challenge classifier achieves the highest accuracy$-$ 99.98\%, with reasonable number of parameters and execution time. Therefore, we select it as our challenge classifier module.

\begin{figure}[!t]
	\centering
	\includegraphics[scale=0.55]{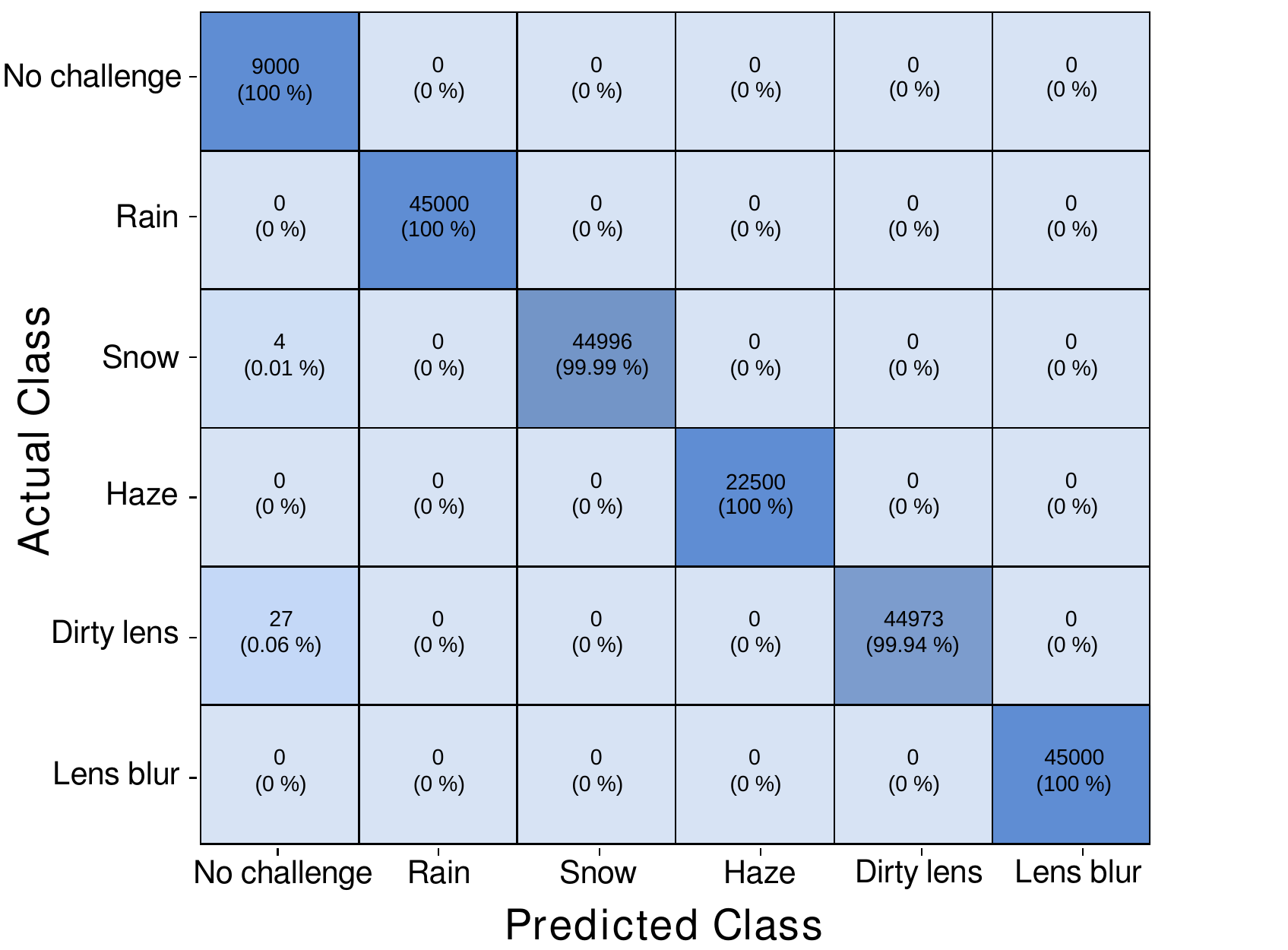}
	\caption{Performance of the challenge classifier.}
	\label{conf_mat}
\end{figure}

One possible phenomenon associated with enhancement of CCs is that, although the performance in CCs improve, performance in challenge-free condition may deteriorate in the case when challenge classifier misclassifies the challenge-free image as a challenging image. In this scenario, a challenge-free image will pass through an enhancement block that can degrade the quality of the image and lead to performance deterioration. However, as we can see from the confusion matrix presented in Fig. \ref{conf_mat}, our challenge classifier achieves 100\% classification accuracy on the CURE-TSD test dataset in detecting challenge-free condition. This rules out the possibility of performance deterioration in challenge-free condition. Misclassification is only encountered in some level $1$ challenges, such as, Snow and Dirty lens, where the challenge classifier misclassifies the conditions as challenge-free condition. From the third and fifth row of the first column of Fig. \ref{fig_challenges}, it is evident that the level $1$ Snow and Dirty Lens challenge are indeed visually very similar to challenge-free condition which makes them comparatively difficult to separate from the challenge-free class.

    
    
         




\begin{table}[h]
\begin{tabular}{|C{2cm}|C{1.5cm}|C{2cm}|C{1.5cm}|}
\hline
Model & Accuracy (\%) & No. of Parameters (Million) & Execution Time (ms)\\
\cline{1-4} 
VGG16     &  99.98 & 14.73 & 3.469\\

\hline
Densenet121     &  99.95 & 6.96 & 2.872\\
     
\hline
Resnet18     &  99.89 & 11.18 & 1.328\\

\hline
\end{tabular}
\caption{Performance Comparison Between Different CNN Architectures as challenge classifier}
\label{classifier_result}
\end{table}

\begin{table*}[!t]
    \centering
    \begin{tabular}{l*{5}{c}r}
    \Xhline{2\arrayrulewidth}
    Model & Data & Loss & Precision & Recall \\
    \Xhline{2\arrayrulewidth}
    
    SegU-Net & challenge-free & $L_{localizer}$ &  83.55   &  34.81  \\

    \hline
    SegU-Net & challenge-free + challenging & $L_{localizer}$ & 80.84 & 39.29 \\
    
    \hline
    SegU-Net + & challenging & $L_{enhancement(overall)}$ &  93.19 & 60.69 \\
    separately trained   & & & & \\
    Enhance-Net   & & & & \\
    
    \hline
    SegU-Net + & challenging & $L_{enhancement(sign)}$ & 92.98 & 64.70 \\
    separately trained   & & & & \\
    Enhance-Net   & & & & \\

    \hline
    SegU-Net + & challenging & $L_{enhancement(overall)}$ &  93.01 & 66.49 \\
    jointly trained   & & + $L_{localizer}$ & & \\
    Enhance-Net   & & & & \\
    
    \hline
    SegU-Net + & challenging & $L_{enhancement(sign)}$ &  91.76 & 68.71 \\
    jointly trained   & & + $L_{localizer}$ & & \\
    Enhance-Net   & & & & \\
    
    \hline
    SegU-Net + & challenging & $L_{enhancement(sign)}$ &  91.68 & 69.89 \\
    jointly trained   & & + $L_{content(sign)}$ & & \\
    Enhance-Net  & & + $L_{localizer}$ & & \\
    (with unity coupling factors)  & & & & \\
    \hline
    SegU-Net + & challenging & $L_{enhancement(sign)}$ &  91.13 & 70.71 \\
    jointly trained   & & + $\lambda_{1} \cdot L_{content(sign)}$ & & \\
    Enhance-Net  & & + $\lambda_{2} \cdot L_{localizer}$ & & \\
    (with optimum coupling factors) & & & & \\
    
    \Xhline{2\arrayrulewidth}
    \end{tabular}
	\caption{Performance on the test set of CURE-TSD dataset with progressive association to the proposed framework}
	\label{full_result_method}
\end{table*}

\subsection{Experiments with Different Approaches}
Next, we select the SegU-Net pretrained on the challenge-free part of the dataset as our sign detection block due to its SOTA performance on the CURE-TSD dataset. In order to validate our motivation mentioned in Section \ref{motivation} and demonstrate the efficacy of the Enhance-Net, we carried out several experiments that are outlined as below:
\begin{enumerate}
    \item SegU-Net re-trained with challenge-free as well as challenging images by using the $L_{localizer}$ loss. No prior enhancement was done on the input image.
	\item Inputs of the SegU-Net are the traffic images enhanced by Enhance-Net. The enhancement block was trained using $L_{enhancement(overall)}$ loss calculated across the whole image. 
	\item Inputs of the SegU-Net are the traffic images enhanced by Enhance-Net. The enhancement block was trained using $L_{enhancement(sign)}$ loss calculated over the sign region only.
    \item Inputs of the SegU-Net are the traffic images enhanced by Enhance-Net. The enhancement block was trained using whole image region-based $L_{enhancement(overall)}$ loss coupled with the $L_{localizer}$ loss. 
    \item Inputs of the SegU-Net are the traffic images enhanced by Enhance-Net. The enhancement block was trained using sign region-based $L_{enhancement(sign)}$ loss coupled with the $L_{localizer}$ loss.
    \item Inputs of the SegU-Net are the traffic images enhanced by Enhance-Net. The enhancement block was trained using sign region-based $L_{enhancement(sign)}$ loss coupled with the $L_{localizer}$ loss and $L_{content(sign)}$ loss with all coupling factors value 1.
    \item Inputs of the SegU-Net are the traffic images enhanced by Enhance-Net. The enhancement block was trained using sign region-based $L_{enhancement(sign)}$ loss coupled with the $L_{localizer}$ loss and $L_{content(sign)}$ loss with the optimum coupling factors.
    
\end{enumerate}

\begin{figure*}[!t]
	\centering
	\includegraphics[width=6.5in, height=3in]{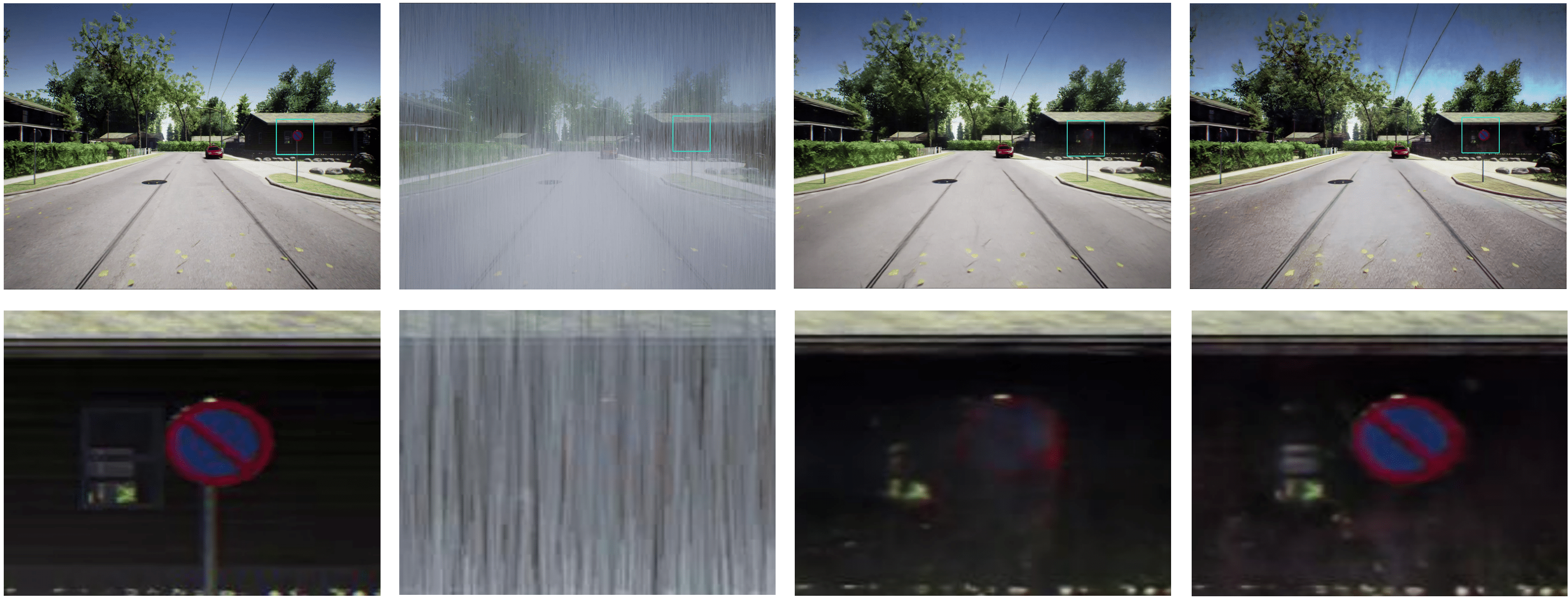}
	\caption{Example of the model output. The first column represents a challenge-free image from CURE-TSD dataset with its zoomed-in sign region. The input challenging (Rain) image and its zoomed version of the sign region are in the second column. The third column shows the output of Enhance-Net trained with overall enhancement-based approach and its corresponding zoomed version of the sign region. The last column shows the output of Enhance-Net for the same input, this time trained with our proposed sign-region-based enhancement approach and its corresponding zoomed version of the sign region.}
	\label{model_output}
\end{figure*} 

\begin{table}[!t]
    \centering
    \begin{tabular}{l*{3}{c}r}
    \Xhline{2\arrayrulewidth}
    Challenge  & SSIM\textsubscript{OE}  & SSIM\textsubscript{SRE} \\
    Levels & (Overall Enhanced) & (Sign Region Enhanced)\\
    \Xhline{2\arrayrulewidth}

    1     &  0.8594   &  0.8750\\
         
    \hline
    2     &  0.8325   &  0.8537 \\

    \hline
    3     &  0.7994   &  0.8258\\

    \hline
    4     &  0.7656   &  0.7972\\

    \hline
    5     &  0.7097   &  0.7465\\

    \hline
    Average &  0.7933   &  0.8196\\

    \Xhline{2\arrayrulewidth}
    \end{tabular}
	\caption{Comparison of Structural Similarity Measure (SSIM) Values between Overall and Only Sign Region Enhancement Approach}
	\label{ssim_compare}
\end{table}

\begin{table}[h]
    \begin{tabular}{|C{2cm}|C{1cm}|C{2cm}|C{2cm}|}
    \hline
    \multirow{2}{*}{Challenge} & \multirow{2}{*}{Level} & \multicolumn{2}{C{4cm}|}{Performance Metrics} \\
    \cline{3-4} 
    
    & & Precision (\%) & Recall (\%) \\
    \Xhline{2\arrayrulewidth}
    No challenge & N/A & 94.60 & 80.21 \\
    \Xhline{2\arrayrulewidth}
    \multirow{5}{*}{Rain}   &  1 &  91.50 & 77.43 \\
    \cline{2-4} 
            &  2 & 90.59 & 74.41 \\
    \cline{2-4} 
            &  3  & 90.24 & 72.26 \\
    \cline{2-4} 
            &  4   &   89.54 & 69.99 \\
    \cline{2-4} 
            &  5   &   87.84 & 63.76 \\
    \Xhline{2\arrayrulewidth}
    \multirow{5}{*}{Snow}   &  1   &   93.81 & 81.30 \\
    \cline{2-4} 
            &  2   &   92.21 & 77.89 \\
    \cline{2-4} 
            &  3   &   93.35 & 72.66 \\
    \cline{2-4} 
            &  4   &   91.22 & 62.01 \\
    \cline{2-4} 
            &  5   &   89.80 & 44.74 \\
    \Xhline{2\arrayrulewidth}
    \multirow{5}{*}{Dirty lens}    &  1   &   93.97 & 81.62 \\
    \cline{2-4} 
                  &  2   &   94.15 & 81.74 \\
    \cline{2-4} 
                  &  3   &   93.87 & 80.61 \\
    \cline{2-4} 
                  &  4   &   93.38 & 79.50 \\
    \cline{2-4} 
                  &  5   &   92.58 & 75.70 \\
    \Xhline{2\arrayrulewidth}
    \multirow{5}{*}{Lens blur}    &  1   &   93.64 & 80.71 \\
    \cline{2-4} 
                 &  2   &   93.03 & 79.85 \\
    \cline{2-4} 
                 &  3   &   93.04 & 78.58 \\
    \cline{2-4} 
                 &  4   &   92.66 & 76.97 \\
    \cline{2-4} 
                 &  5   &   92.64 & 73.82 \\
    \Xhline{2\arrayrulewidth}
    \multirow{5}{*}{Haze}    &  1   &   86.81 & 58.96 \\
    \cline{2-4} 
            &  2   &   87.02 & 58.70 \\
    \cline{2-4} 
            &  3   &   86.94 & 57.82 \\
    \cline{2-4} 
            &  4   &   87.07 & 56.12 \\
    \cline{2-4} 
            &  5   &   87.30 & 50.66 \\






    \Xhline{2\arrayrulewidth}
    \end{tabular}
	\caption{Results of the proposed method on the test set of CURE-TSD Dataset for different challenging conditions}
	\label{full_result}
\end{table}

\begin{table}[!t]
    \centering
    \begin{tabular}{l*{5}{c}r}
    \Xhline{2\arrayrulewidth}
    Method  & Precision  & Recall \\
    \Xhline{2\arrayrulewidth}
    R-FCN     &  53.31   &  44.48 \\
    \hline
    Faster-RCNN     &  58.17   &  46.03 \\
    \hline
    Proposed Method     &  \textbf{91.13}   &  \textbf{70.71} \\

    \Xhline{2\arrayrulewidth}
    \end{tabular}
	\caption{Comparison with end to end trained Frameworks}
	\label{framework_compare}
\end{table}



The experimental results are summarized in Table \ref{full_result_method}. From the first result in this table, we observe that training the sign detection module with abundant training data is not effective in alleviating performance degradation caused by the challenging conditions. This justifies our motivation for using the enhancement module. From the other results, we can infer that enhancement of traffic sign regions should be the main focus of the enhancement module for better traffic sign detection performance. To further confirm this, we present the Structural Similarity Measure (SSIM) between the enhanced and challenge-free traffic sign regions for both overall and only sign region enhancement approach. Table \ref{ssim_compare} shows the challenges severity level-wise SSIM values, averaged over all the challenge types. From this table, we see that our proposed method achieves higher SSIM score$-$ $0.8196$ across the traffic sign regions compared to the $0.7933$ SSIM score from overall enhancement approach. It is also evident that there is a degradation of SSIM values associated with the increasing level of difficulty for both cases. However, our proposed approach suffers from less performance deterioration than the conventional approach. We also show some outputs of our proposed method and the outputs of the overall enhanced method side by side in Fig. \ref{model_output}. From this figure, we can see that although the overall image achieves better visibility in the case of total enhancement approach, the sign region still suffers from harsh visibility. Whereas, in our approach, the sign region achieves almost similar visibility compared to the challenge-free condition. Finally, the incorporation of $L_{localizer}$ brings forth further performance gain in the overall approach because it constrains the training of Enhance-Net to learn to enhance the sign regions subject to their accurate detection by the subsequent sign detection module. It is noteworthy to mention that, overall enhancement-based approaches achieve higher precision values which can be observed from the third and fifth row of Table \ref{full_result_method}. It is due to the fact that, overall enhancement based approach ensures better image quality for the background region, which can assist the sign detector module in the false positive reduction. However, if we consider both precision and recall, our approach achieves better overall performance. 

\subsection{Performance on CURE-TSD}
\begin{figure}[!t]
	\centering
	\includegraphics[width=3.5in, height=1.6in]{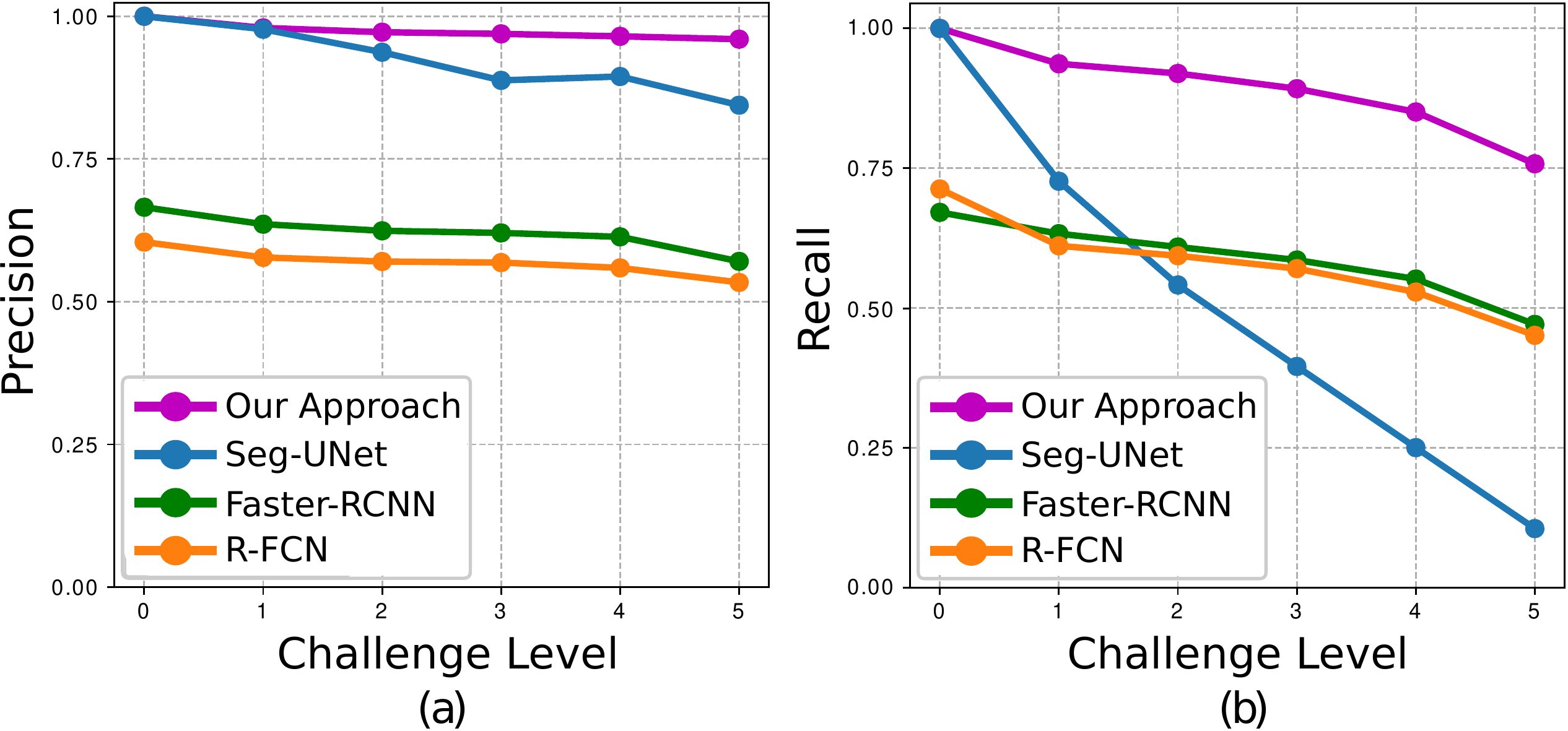}
	\caption{Performance degradation of different methods on CURE-TSD dataset with the increasing difficulty of the challenges.}
	\label{graph_of_performance}
\end{figure}

We evaluate the performance of our proposed method on the selected set of the challenging part of CURE-TSD. We report the performance on the challenge-free type as well as the five selected challenge types from the test set of CURE-TSD dataset in Table \ref{full_result}. For `No-Challenge' type, the precision and recall$-$ $94.60\%$ and $80.21\%$ are as same as reported in \cite{uday} because we use their pretrained model on challenge-free part of the dataset. Results on different challenge types highlight the fact that there is still some performance degradation associated with the higher level of CCs, specially for level 4 and 5. These higher levels of challenge create an extremely harsh visibility condition that can be observed in Fig. \ref{fig_challenges} as well. For instance, from the fifth row of the fifth column of Fig. \ref{fig_challenges}, i.e. level 5 Snow challenge, it is evident that the sign region has almost completely lost all its shape and color information that adversely affects the performance of the sign detector. As a result, this particular case has the least recall of $44.74\%$ among all other cases. However, we want to emphasize the fact that a carefully designed enhancement architecture might be able to alleviate this problem, which we outline as the future research direction. Also, for Haze, the overall performance metrics are relatively lower compared to the other CCs. This can be explained by the lack of sufficient data for this case. As we mentioned in section \ref{dataset}, for this specific condition, there was almost half the amount of training data compared to other CCs. It is due to the absence of a simulated environment for this specific type of challenge alone. However, we believe that with sufficient data, the performance on Haze can be further improved. To demonstrate the superiority of our modular structure, we compare its performance to the two different object detection networks$-$ Faster R-CNN and R-FCN with Inception Resnet V2 and Resnet 101 as backbone respectively, which are reported to achieve SOTA traffic sign detection performance on GTSDB dataset \cite{arcos2018evaluation}. Kamal \textit{et. al} \cite{uday} have also used these networks to compare their method's performance on the challenge-free part of CURE-TSD dataset. In this work, we train them on both challenging and challenge-free parts of the dataset. For training and testing purpose, we use the TensorFlow Object Detection API where the networks were initialized with weights pretrained on the Microsoft COCO dataset \cite{lin2014microsoft}. Table \ref{framework_compare} contains the comparative results obtained by these methods and our approach. Here, the Precision and Recall scores are averaged over all five levels of our selected challenge types along with the challenge-free part of the dataset. Our approach obtains overall $91.13\%$ and  $70.71\%$ precision and recall respectively, which outperforms the results obtained by Faster-RCNN$-$ $58.17\%$, $46.03\%$ and R-FCN$-$ $53.31\%$, $44.48\%$ precision and recall, respectively, by a large margin. This suggests the fact that, without special attention to the challenge types, these networks trained with a large volume of data in end-to-end fashion can not achieve competitive performance. On the other hand, the modular structure of our approach has specialized networks for each stage of the TSDR task that also includes challenge specific prior enhancement. This process ensures the best achievable performance on each task, i.e. enhancement, sign detection, and sign recognition, independently, which results in an overall superior performance in terms of both precision and recall. We also show the comparison of the performance degradation associated with the increasing challenge levels for different approaches in Fig. \ref{graph_of_performance}. It is evident that the precision score does not deteriorate much compare to the recall score for all four approaches because the number of false detection does not increase much with the increasing challenge levels. However, this is not the case for recall score. With increasing challenge levels, the sign regions become more obscured which makes them very hard to detect. Among the four approaches, Seg-UNet, trained on the challenge free dataset only, suffers from the highest performance degradation due to the lack of challenging train dataset. If we compare the performance of the deeper CNN models, i.e. Faster-RCNN and R-FCN, trained on both challenging and challenge-free dataset, we can see that although they achieve lower detection rate on challenge-free condition, their performance degradation rate is under a tolerable margin. For these two methods, training with both challenging and challenge-free dataset helped to reduce the performance degradation rate. However, without any special attention for the challenge parts, their method fails to achieve the best performance. This justifies our initial hypothesis that end-to-end training with abundant data is not sufficient to achieve the desired result for this task. On the other hand, our modular approach integrated with separate enhancement blocks for each challenge achieves the least performance (both precision and recall) degradation while maintaining the baseline performance for the challenge-free type as well. This experimentally proves the superiority of our enhancement-based modular approach over the existing SOTA methods for robust traffic sign detection even under severe CCs.

\section{Conclusion}
In this paper, we have presented a deep CNN-based modular
and a robust framework for TSDR under various CCs. We have
highlighted the performance degradation of the existing TSDR
algorithms due to the presence of different CCs and proposed
a deep CNN-based approach that effectively alleviates the
problem. A VGG16 architecture-based challenge classifier, that
successfully detects and classifies the challenge, directs the image to the appropriate Enhance-Net which recovers the features that are useful for the successful detection of the traffic sign regions. Unlike the existing whole image enhancement-based methods, the Enhance-Nets are trained by
our proposed novel loss function and training pipeline that
incorporate traffic sign region focused MAE in both pixel and
feature domain with the sign detection loss as a constraint.
This effectively ensures the enhancement of the sign regions
subject to their accurate detection. We have also experimentally showed that traffic sign regions are more important for
enhancement, in order to obtain higher detection performance.
Finally, we evaluate the efficacy of the modular structure of
our approach by comparing its performance with two different
end-to-end trained deep CNN-based object detection networks
where our approach outperforms both of them. Due to our
modular approach, each module of our framework can be
designed independently. This opens up a vast research scope
in this field.
As future work, we wish to explore and design the
optimum architecture for each module so that all of the CCs
present in the CURE-TSD dataset are best addressed.

\bibliographystyle{IEEEtran}
\bibliography{references}

\begin{thebibliography}{10}
\providecommand{\url}[1]{#1}
\csname url@samestyle\endcsname
\providecommand{\newblock}{\relax}
\providecommand{\bibinfo}[2]{#2}
\providecommand{\BIBentrySTDinterwordspacing}{\spaceskip=0pt\relax}
\providecommand{\BIBentryALTinterwordstretchfactor}{4}
\providecommand{\BIBentryALTinterwordspacing}{\spaceskip=\fontdimen2\font plus
\BIBentryALTinterwordstretchfactor\fontdimen3\font minus
  \fontdimen4\font\relax}
\providecommand{\BIBforeignlanguage}[2]{{%
\expandafter\ifx\csname l@#1\endcsname\relax
\typeout{** WARNING: IEEEtran.bst: No hyphenation pattern has been}%
\typeout{** loaded for the language `#1'. Using the pattern for}%
\typeout{** the default language instead.}%
\else
\language=\csname l@#1\endcsname
\fi
#2}}
\providecommand{\BIBdecl}{\relax}
\BIBdecl

\bibitem{mogelmose2012vision}
A.~Mogelmose, M.~M. Trivedi, and T.~B. Moeslund, ``Vision-based traffic sign
  detection and analysis for intelligent driver assistance systems:
  Perspectives and survey,'' \emph{IEEE Transactions on Intelligent
  Transportation Systems}, vol.~13, no.~4, pp. 1484--1497, 2012.

\bibitem{gudigar2016review}
A.~Gudigar, S.~Chokkadi, and U.~Raghavendra, ``A review on automatic detection
  and recognition of traffic sign,'' \emph{Multimedia Tools and Applications},
  vol.~75, no.~1, pp. 333--364, 2016.

\bibitem{xu2009robust}
S.~Xu, ``Robust traffic sign shape recognition using geometric matching,''
  \emph{IET Intelligent Transport Systems}, vol.~3, no.~1, pp. 10--18, 2009.

\bibitem{khan2011image}
J.~F. Khan, S.~M. Bhuiyan, and R.~R. Adhami, ``Image segmentation and shape
  analysis for road-sign detection,'' \emph{IEEE Transactions on Intelligent
  Transportation Systems}, vol.~12, no.~1, pp. 83--96, 2011.

\bibitem{creusen2010color}
I.~M. Creusen, R.~G. Wijnhoven, E.~Herbschleb, and P.~de~With, ``Color
  exploitation in hog-based traffic sign detection,'' in \emph{Image Processing
  (ICIP), 2010 17th IEEE International Conference on}.\hskip 1em plus 0.5em
  minus 0.4em\relax IEEE, 2010, pp. 2669--2672.

\bibitem{temel_2}
D.~Temel, M.-H. Chen, and G.~AlRegib, ``Traffic sign detection under
  challenging conditions: A deeper look into performance variations and
  spectral characteristics,'' in \emph{IEEE Transactions on Intelligent
  Transportation Systems}, 2019.

\bibitem{canny1987computational}
J.~Canny, ``A computational approach to edge detection,'' in \emph{Readings in
  Computer Vision}.\hskip 1em plus 0.5em minus 0.4em\relax Elsevier, 1987, pp.
  184--203.

\bibitem{baro2009traffic}
X.~Bar{\'o}, S.~Escalera, J.~Vitri{\`a}, O.~Pujol, and P.~Radeva, ``Traffic
  sign recognition using evolutionary adaboost detection and forest-ecoc
  classification,'' \emph{IEEE Transactions on Intelligent Transportation
  Systems}, vol.~10, no.~1, pp. 113--126, 2009.

\bibitem{jimenez2008traffic}
P.~G. Jim{\'e}nez, S.~M. Basc{\'o}n, H.~G. Moreno, S.~L. Arroyo, and F.~L.
  Ferreras, ``Traffic sign shape classification and localization based on the
  normalized fft of the signature of blobs and 2d homographies,'' \emph{Signal
  Processing}, vol.~88, no.~12, pp. 2943--2955, 2008.

\bibitem{zaklouta2014real}
F.~Zaklouta and B.~Stanciulescu, ``Real-time traffic sign recognition in three
  stages,'' \emph{Robotics and autonomous systems}, vol.~62, no.~1, pp. 16--24,
  2014.

\bibitem{krizhevsky2012ImageNet}
A.~Krizhevsky, I.~Sutskever, and G.~E. Hinton, ``Imagenet classification with
  deep convolutional neural networks,'' in \emph{Advances in neural information
  processing systems}, 2012, pp. 1097--1105.

\bibitem{chen2014semantic}
L.-C. Chen, G.~Papandreou, I.~Kokkinos, K.~Murphy, and A.~L. Yuille, ``Semantic
  image segmentation with deep convolutional nets and fully connected crfs,''
  \emph{arXiv preprint arXiv:1412.7062}, 2014.

\bibitem{girshick2014rich}
R.~Girshick, J.~Donahue, T.~Darrell, and J.~Malik, ``Rich feature hierarchies
  for accurate object detection and semantic segmentation,'' in
  \emph{Proceedings of the IEEE conference on computer vision and pattern
  recognition}, 2014, pp. 580--587.

\bibitem{conv1}
H.~{Luo}, Y.~{Yang}, B.~{Tong}, F.~{Wu}, and B.~{Fan}, ``Traffic sign
  recognition using a multi-task convolutional neural network,'' \emph{IEEE
  Transactions on Intelligent Transportation Systems}, vol.~19, no.~4, pp.
  1100--1111, April 2018.

\bibitem{conv2}
Y.~{Yang}, H.~{Luo}, H.~{Xu}, and F.~{Wu}, ``Towards real-time traffic sign
  detection and classification,'' \emph{IEEE Transactions on Intelligent
  Transportation Systems}, vol.~17, no.~7, pp. 2022--2031, July 2016.

\bibitem{frcnn_mobilenet}
J.~{Li} and Z.~{Wang}, ``Real-time traffic sign recognition based on efficient
  cnns in the wild,'' \emph{IEEE Transactions on Intelligent Transportation
  Systems}, vol.~20, no.~3, pp. 975--984, March 2019.

\bibitem{lee}
H.~S. {Lee} and K.~{Kim}, ``Simultaneous traffic sign detection and boundary
  estimation using convolutional neural network,'' \emph{IEEE Transactions on
  Intelligent Transportation Systems}, vol.~19, no.~5, pp. 1652--1663, May
  2018.

\bibitem{vssa}
Y.~{Yuan}, Z.~{Xiong}, and Q.~{Wang}, ``Vssa-net: Vertical spatial sequence
  attention network for traffic sign detection,'' \emph{IEEE Transactions on
  Image Processing}, vol.~28, no.~7, pp. 3423--3434, July 2019.

\bibitem{houben2013detection}
S.~Houben, J.~Stallkamp, J.~Salmen, M.~Schlipsing, and C.~Igel, ``Detection of
  traffic signs in real-world images: The german traffic sign detection
  benchmark,'' in \emph{Neural Networks (IJCNN), The 2013 International Joint
  Conference on}.\hskip 1em plus 0.5em minus 0.4em\relax IEEE, 2013, pp. 1--8.

\bibitem{lisa}
A.~{Mogelmose}, M.~M. {Trivedi}, and T.~B. {Moeslund}, ``Vision-based traffic
  sign detection and analysis for intelligent driver assistance systems:
  Perspectives and survey,'' \emph{IEEE Transactions on Intelligent
  Transportation Systems}, vol.~13, no.~4, pp. 1484--1497, Dec 2012.

\bibitem{timofte2014multi}
R.~Timofte, K.~Zimmermann, and L.~Van~Gool, ``Multi-view traffic sign
  detection, recognition, and 3d localisation,'' \emph{Machine vision and
  applications}, vol.~25, no.~3, pp. 633--647, 2014.

\bibitem{temel2017cure-tsd}
D.~Temel, T.~Alshawi, M.-H. Chen, and G.~AlRegib, ``Cure-tsd: Challenging
  unreal and real environments for traffic sign detection,'' August 2017,
  submitted to IEEE Transactions on Intelligent Transportation Systems.

\bibitem{uday}
U.~{Kamal}, T.~I. {Tonmoy}, S.~{Das}, and M.~K. {Hasan}, ``Automatic traffic
  sign detection and recognition using segu-net and a modified tversky loss
  function with l1-constraint,'' \emph{IEEE Transactions on Intelligent
  Transportation Systems}, pp. 1--13, 2019.

\bibitem{derain}
Y.~{Li}, R.~T. {Tan}, X.~{Guo}, J.~{Lu}, and M.~S. {Brown}, ``Single image rain
  streak decomposition using layer priors,'' \emph{IEEE Transactions on Image
  Processing}, vol.~26, no.~8, pp. 3874--3885, Aug 2017.

\bibitem{derain_2}
Y.~{Luo}, J.~{Zhu}, J.~{Ling}, and E.~{Wu}, ``Fast removal of rain streaks from
  a single image via a shape prior,'' \emph{IEEE Access}, vol.~6, pp.
  60\,069--60\,078, 2018.

\bibitem{dehazing}
J.~Li, G.~Li, and H.~Fan, ``Image dehazing using residual-based deep cnn,''
  \emph{IEEE Access}, vol.~PP, pp. 1--1, 05 2018.

\bibitem{rain_and_snow}
J.~{Kim}, J.~{Sim}, and C.~{Kim}, ``Video deraining and desnowing using
  temporal correlation and low-rank matrix completion,'' \emph{IEEE
  Transactions on Image Processing}, vol.~24, no.~9, pp. 2658--2670, Sep. 2015.

\bibitem{rain_and_snow_1}
W.~{Ren}, J.~{Tian}, Z.~{Han}, A.~{Chan}, and Y.~{Tang}, ``Video desnowing and
  deraining based on matrix decomposition,'' in \emph{2017 IEEE Conference on
  Computer Vision and Pattern Recognition (CVPR)}, July 2017, pp. 2838--2847.

\bibitem{srgan}
C.~{Ledig}, L.~{Theis}, F.~{Huszár}, J.~{Caballero}, A.~{Cunningham},
  A.~{Acosta}, A.~{Aitken}, A.~{Tejani}, J.~{Totz}, Z.~{Wang}, and W.~{Shi},
  ``Photo-realistic single image super-resolution using a generative
  adversarial network,'' in \emph{2017 IEEE Conference on Computer Vision and
  Pattern Recognition (CVPR)}, July 2017, pp. 105--114.

\bibitem{arcos2018evaluation}
{\'A}.~Arcos-Garc{\'\i}a, J.~A. {\'A}lvarez-Garc{\'\i}a, and L.~M.
  Soria-Morillo, ``Evaluation of deep neural networks for traffic sign
  detection systems,'' \emph{Neurocomputing}, vol. 316, pp. 332--344, 2018.

\bibitem{vgg}
K.~Simonyan and A.~Zisserman, ``Very deep convolutional networks for
  large-scale image recognition,'' \emph{arXiv 1409.1556}, 09 2014.

\bibitem{long2015fully}
J.~Long, E.~Shelhamer, and T.~Darrell, ``Fully convolutional networks for
  semantic segmentation,'' in \emph{Proceedings of the IEEE conference on
  computer vision and pattern recognition}, 2015, pp. 3431--3440.

\bibitem{dgan}
O.~{Kupyn}, V.~{Budzan}, M.~{Mykhailych}, D.~{Mishkin}, and J.~{Matas},
  ``Deblurgan: Blind motion deblurring using conditional adversarial
  networks,'' in \emph{2018 IEEE/CVF Conference on Computer Vision and Pattern
  Recognition}, June 2018, pp. 8183--8192.

\bibitem{resnet}
K.~{He}, X.~{Zhang}, S.~{Ren}, and J.~{Sun}, ``Deep residual learning for image
  recognition,'' in \emph{2016 IEEE Conference on Computer Vision and Pattern
  Recognition (CVPR)}, June 2016, pp. 770--778.

\bibitem{kingma2014adam}
D.~P. Kingma and J.~Ba, ``Adam: A method for stochastic optimization,''
  \emph{arXiv preprint arXiv:1412.6980}, 2014.

\bibitem{l2}
M.~Mathieu, C.~Couprie, and Y.~Lecun, ``Deep multi-scale video prediction
  beyond mean square error,'' 2016.

\bibitem{l1_better}
H.~Zhao, O.~Gallo, I.~Frosio, and J.~Kautz, ``Is l2 a good loss function for
  neural networks for image processing?'' \emph{arXiv preprint
  arXiv:1511.08861}, 2015.

\bibitem{cnn_features}
M.~Zeiler and R.~Fergus, ``Visualizing and understanding convolutional neural
  networks,'' vol. 8689, 11 2013.

\bibitem{edge_loss}
H.~{Zhang} and V.~M. {Patel}, ``Densely connected pyramid dehazing network,''
  in \emph{2018 IEEE/CVF Conference on Computer Vision and Pattern
  Recognition}, June 2018, pp. 3194--3203.

\bibitem{lin2014microsoft}
T.-Y. Lin, M.~Maire, S.~Belongie, J.~Hays, P.~Perona, D.~Ramanan,
  P.~Doll{\'a}r, and C.~L. Zitnick, ``Microsoft coco: Common objects in
  context,'' in \emph{European conference on computer vision}.\hskip 1em plus
  0.5em minus 0.4em\relax Springer, 2014, pp. 740--755.

\end{thebibliography}

\end{document}